\documentclass[a4paper,fleqn,usenatbib]{mnras}
\pdfoutput=1
\usepackage{newtxtext,newtxmath}

\usepackage[T1]{fontenc}
\usepackage{ae,aecompl}

\usepackage{graphicx}	
\usepackage{amsmath}	
\usepackage{amssymb}	

\usepackage{multirow}
\usepackage{soul}

\graphicspath{{./figs/}}

\title[Properties of discs with clumpy episodes]{Geometric properties
  of galactic discs with clumpy episodes}

\author[L. {Beraldo e Silva} et al.]{
Leandro {Beraldo e Silva}$^{1}$\thanks{E-mail: lberaldoesilva@uclan.ac.uk (LBS)},
Victor P. Debattista$^{1}$,
Tigran Khachaturyants$^{1}$,
David Nidever$^{2,3}$
\\
$^{1}$Jeremiah Horrocks Institute, University of Central Lancashire,
Preston, PR1 2HE, UK\\
$^{2}$Department of Physics, Montana State University, P.O. Box
173840, Bozeman, MT 59717-3840, USA\\
$^{3}$National Optical Astronomy Observatory, 950 North Cherry Ave, Tucson, AZ 85719, USA
}

\date{Accepted XXX. Received
  YYY; in original form ZZZ}

\pubyear{2019}

\begin{document}
\label{firstpage}
\pagerange{\pageref{firstpage}--\pageref{lastpage}}
\maketitle

\begin{abstract}
  A scenario for the formation of the bi-modality in the chemical
  space [$\alpha$/Fe] vs [Fe/H] of the Milky Way was recently proposed
  in which $\alpha$-enhanced stars are produced early and quickly in
  clumps. Besides accelerating the enrichment of the medium with
  $\alpha$-elements, these clumps scatter the old stars, converting
  in-plane to vertical motion, forming a geometric thick disc. In this
  paper, by means of a detailed analysis of the data from smooth
  particle hydrodynamical simulations, we investigate the geometric
  properties (in particular of the chemical thick disc) produced in
  this scenario. For mono-age populations we show that the surface
  radial density profiles of high-[$\alpha$/Fe] stars are well
  described by single exponentials, while that of low-[$\alpha$/Fe]
  stars require broken exponentials. This break is sharp for young
  populations and broadens for older ones. The position of the break
  does not depend significantly on age. The vertical density profiles
  of mono-age populations are characterized by single exponentials,
  which flare significantly for low-[$\alpha$/Fe] stars but only
  weakly (or not at all) for high-[$\alpha$/Fe] stars. For
  low-[$\alpha$/Fe] stars, the flaring level decreases with age, while
  for high-[$\alpha$/Fe] stars it weakly increases with age (although
  with large uncertainties). All these properties are in agreement
  with observational results recently reported for the Milky Way,
  making this a plausible scenario for the formation of the Galactic
  thick disc.
\end{abstract}




\section{Introduction}
\label{sec:intro}
Galactic discs have two distinct photometric (or geometric)
components, the thin and the thick discs, as first discovered by
\cite{1979ApJ...234..829B} and \cite{1979ApJ...234..842T} in external
galaxies and further explored in more systematic studies by
e.g. \cite{2006AJ....131..226Y,
  2008ApJ...682.1004Y,2008ApJ...683..707Y} in the local universe and
by \cite{2006ApJ...650..644E,2017ApJ...847...14E} at high
redshifts. The same composite structure was also observed in the Milky
Way (MW) since \cite{1982PASJ...34..365Y} and
\cite{1983MNRAS.202.1025G}, with more recent results showing that in
the Solar neighborhood the thin disc is mainly composed of younger
stars which are poor in $\alpha$-elements and metal-rich, and with a
characteristic scale height $h_{z1}\approx 0.3$ kpc. On the other
hand, the thick disc is characterized by stars that are old,
$\alpha$-enhanced and metal-poor, with vertical scale height
$h_{z2}\approx 0.9$ kpc -- see
e.g. \cite{2005A&A...433..185B,2008ApJ...673..864J}.

The origin of the thick disc is still under active debate. Some works
suggest the ``upside-down'' formation scenario, in which stars in the
thick disc form while the gas is collapsing, i.e. they already form
with their current height distribution. In a different scenario,
\cite{2009ApJ...707L...1B} observed that gas-rich young galaxies in
simulations produce clumps that strongly scatter old stars, converting
in-plane to vertical motion and giving origin to a thick
disc. \cite{2009ApJ...707L...1B} also concluded that a thick disc
formed in this way has a constant scale height with galactocentric
radius, i.e. it does not flare. Some observational support for this
scenario was obtained by
\cite{2011ApJ...741...28C,2014A&A...571A..58C}. Another interesting
possibility for the formation of the thick disc is that these stars
were brought (or produced in situ) by a major merger around 10 Gyr ago
-- see e.g.  \cite{2018MNRAS.478..611B, 2018Natur.563...85H}. The
radial migration (of stars at corotation with transient spiral arms)
was also proposed as a mechanism for the formation of the thick disc
\citep[][]{2009MNRAS.399.1145S, 2011ApJ...737....8L}, although it
seems hard to reconcile this scenario with the bi-modality in the
$\alpha$-abundance observed in the Milky Way (see below).  On the
other hand, radial migration is generally seen as a fundamental
process determining the stellar radial distribution in the thin disc,
in particular shaping the metallicity distribution function at
different radii -- see \cite{2015ApJ...808..132H,
  2016ApJ...818L...6L}.

Selecting mono-abundance populations (MAPs) in the chemical space
[$\alpha$/Fe] vs [Fe/H], \cite{2012ApJ...753..148B} concluded that the
vertical density profiles of each of these MAPs in the MW is
characterized by only one scale height, i.e. by a single exponential,
finding no evidence of a chemically distinct thick disc and that the
thin + thick disc structure only emerges as a consequence of
superposing populations with different physical properties and
dynamical histories. Similar conclusions were reached by
\cite{2015ApJ...804L...9M}, who propose that the geometric thick disc
is a consequence of the superposition of many subpopulations, which
are well fit by single exponentials with different flaring levels.

Closely related to the existence of the geometric thin and thick discs
is the presence of a bi-modality in chemical space ([$\alpha$/Fe] vs
[Fe/H]), defining a high-$\alpha$ and a low-$\alpha$ disc, as revealed
by the APOGEE survey
\citep[][]{2014AA...564A.115A,2014ApJ...796...38N,2015ApJ...808..132H},
showing the existence of a chemically distinct thick disc (the
high-$\alpha$ stars), while the low-$\alpha$ region defines the
chemical thin disc, although the association with the geometric discs
is not exact. Some studies have associated this chemical bi-modality
to the accretion events of the Galaxy \citep[see
e.g.][]{10.1093/mnras/sty972,2020MNRAS.491.5435B}. However, its origin
is still open.

Recently, \cite{2019MNRAS.484.3476C} proposed that this bi-modality is
associated to the occurrence of two distinct star formation channels
in the disc: an early and fast star formation with fast chemical
enrichment in clumps and a more continuous star formation dispersed in
the disc. In this scenario, the high star formation rate density in
clumps generates many supernovae type II, which rapidly enrich the
medium with $\alpha$ elements. Additionally, as in the scattering
process by giant molecular clouds proposed by
\cite{1953ApJ...118..106S}, the clumps scatter the old, $\alpha$-rich
stars, converting in-plane to vertical motion and promoting stars to
high latitudes \citep[essentially the same mechanism proposed
by][]{2009ApJ...707L...1B}. \cite{2019MNRAS.484.3476C} ran
hydrodynamical simulations employing a low stellar feedback
efficiency, allowing the formation of clumps. They showed that these
simulations naturally produce a chemical bi-modality with properties
similar to the ones observed in the MW - see Fig. 5 from
\cite{2019MNRAS.484.3476C}.

In the current work, we investigate the spatial distribution of the
chemical discs formed by this simulation. In particular, we test
whether, besides naturally producing the chemical bi-modality, the
simple scenario explored in this simulation is also able to produce
realistic geometric discs, particularly the thick disc. We apply
several different cuts and compare the resulting density profiles with
observational results. In Sec. \ref{sec:sims}, we describe the
simulation. In Sec. \ref{sec:models}, we introduce the different
fitting models, which are compared to the simulation data in
Sec.\ref{sec:results}. In Sec. \ref{sec:compare_results}, we compare
our results with previous observational and simulation-based results
and finally in Sec.\ref{sec:conclusions}, we discuss and summarize the
conclusions.

\section{The Simulations}
\label{sec:sims}

The simulation we use in this paper is the same as that explored and
described by \cite{2019MNRAS.484.3476C}. Briefly, the model starts
with a hot gas corona embedded in a spherical dark matter halo with a
Navarro-Frenk-White \citep[][]{1997ApJ...490..493N} halo with virial
radius $r_{200} \simeq 200$ kpc and mass of $10^{12} M_\odot$. The gas
has spin $\lambda = 0.065$ \citep{2001ApJ...555..240B}. The dark
matter halo and gas corona both consist of $10^6$ particles
initially. The gas cools via metal-line cooling
\citep{2010MNRAS.407.1581S}, and settles into a disc, with stars
forming from this gas when the temperature drops below 15,000~K and
the density exceeds $1 \mathrm{cm}^{-3}$. Feedback via supernova
explosions uses the blastwave implementation described in
\citet{2006MNRAS.373.1074S}, with 10\% of the $10^{51}$ erg per
supernova injected to the interstellar medium as thermal energy.
Feedback via asymptotic giant branch stars is also taken into account.
Following \citet{2009MNRAS.397L..64A}, we prevent gas cooling from
dropping below our resolution by setting a gas pressure floor of
$p_{floor} = 3 G\epsilon^2\rho^2$, where $G$ is Newton's gravitational
constant, $\epsilon$ is the softening length set at $50$ pc and $\rho$
is the gas particle's density.  Gas phase diffusion uses the method of
\citet{2010MNRAS.407.1581S}, reducing the scatter in the
age-metallicity relation \citep{2012MNRAS.425..969P}.

We evolve the model using the smooth particle hydrodynamics$+N$-body
tree-code {\sc gasoline} \citep{gasoline}. The metal-line cooling
results in clumps forming during early times in the model, as shown in
\cite{2019MNRAS.484.3476C}. The final disc galaxy that forms has a
rotational velocity of $242$ km/s at the Solar neighborhood, making it
comparable to the MW.  The final rotation curve and velocity
dispersion profiles of the model are shown in Figure 2 of
\cite{2019MNRAS.484.3476C}. We will refer to this simulation as {\bf
  FB10} (for feedback efficiency of $10\%$). Note that, although this
simulated galaxy presents realistic properties in general, it is not
intended to be a replica of the Milky Way. In particular, this
simulated galaxy does not form a bar and has spatial scales with
absolute values differing from those of the Milky Way.

In order to illustrate the effect of clumps on the chemistry of the
model, we compare with a model in which the feedback from supernovae
is set at $80\%$ of the $10^{51}$ erg per supernova. This feedback is
8 times higher than in our fiducial simulation; as a result clump
formation is significantly suppressed in this simulation, as has been
found in previous simulations \citep{hopkins+12, genel+12, buck+17,
  oklopcic+17}. We refer to this model as the high-feedback model (or
{\bf FB80}).

\section{Modelling the Density profiles}
\label{sec:models}
In order to analyze the simulation data and make comparisons with
observational results, we explore several fitting functions inspired
by the models used in observational analyses. Our modelling varies in
complexity reflecting approximately the chronological order in which
these models were proposed by different studies. In all these models,
the star-count density profiles are generically written as
\begin{equation}
  \label{eq:1}
  \nu_\star(R,z|\theta) = \Sigma(R|\theta)\zeta(z|R,\theta),
\end{equation}
where $R$ is the cylindrical radius, $z$ is the absolute latitude
(vertical distance above or below the galactic plane), $\Sigma(R)$ is
the surface density (units of length$^{-2}$), $\zeta(z|R)$ is the
vertical density profile (units of length$^{-1}$) and $\theta$ is the
set of parameters of the model.

We consider a total of five models (see Table \ref{tab:models}), among
which the first four have a a single exponential surface density
profile , i.e.
\begin{equation}
  \label{eq:Sigma_single_exp}
  \Sigma(R) = \frac{1}{A}e^{-R/h_R},
\end{equation}
where $h_R$ is a free parameter and the constant $A$ is determined by
the normalization condition
\begin{equation}
  \label{eq:normal}
  \int_{z_{min}}^{z_{max}}dz\int_{R_{min}}^{R_{max}}\nu_\star(R,z|\theta)\, 2\pi RdR
  = N_\star,
\end{equation}
where $N_\star$ is the total number of star particles.

The first model (${\bf M1}$) considers a double exponential for the
vertical profile, defining the thin and thick discs as first
identified in the MW by \cite{1982PASJ...34..365Y} and
\cite{1983MNRAS.202.1025G}:
\begin{equation}
  \label{eq:rho_double_exp}
  \zeta(z|R) = \frac{1-\beta}{2h_{z1}}e^{-|z|/h_{z1}} + \frac{\beta}{2h_{z2}}e^{-|z|/h_{z2}},
\end{equation}
where $h_{z1}$, $h_{z2}$ and $\beta$ are also free parameters.

Inspired by the results of \cite{2012ApJ...753..148B}, who found that
the vertical density profile of mono-abundance populations in the MW
were better described by single exponentials, our second model
(${\bf M2}$) is characterized by
\begin{equation}
  \label{eq:rho_single_exp}
  \zeta(z|R) = \frac{1}{2h_z}e^{-|z|/h_z}.
\end{equation}
In Eqs.~(\ref{eq:rho_double_exp}) and (\ref{eq:rho_single_exp}), the
$R$-dependence is only formally present, since in these models the
parameters $h_z$, $\beta$, $h_{z1}$ and $h_{z2}$ do not depend on $R$.

In our third model (${\bf M3}$), following the parameterization of
\cite{2016ApJ...823...30B}, the vertical density profile is still a
single exponential, Eq.~(\ref{eq:rho_single_exp}), but now the
vertical scale height is allowed to vary with radius $R$, i.e. to
flare, as
\begin{equation}
  \label{eq:hz_flare}
  h_z(R) = h_{z\odot}e^{R_{fl}^{-1}(R-R_\odot)},
\end{equation}
where the characteristic flaring radius $R_{fl}$ and the scale height
at the Solar position $h_{z\odot}$ are also free
parameters\footnote{Note that our definition of $R_{fl}$ differs from
  that of \cite{2016ApJ...823...30B,2017MNRAS.471.3057M} by a minus
  sign.}, while $R_\odot=8$ kpc is the radius of Solar position. Model
${\bf M4}$ is characterized by a double exponential vertical density
profile, Eq.~(\ref{eq:rho_double_exp}), but now the inner vertical
scale height $h_{z1}$ is allowed to flare as in
Eq.~(\ref{eq:hz_flare}).

Finally, in model ${\bf M5}$ the vertical density profile is modeled
as a single flaring exponential (as in model ${\bf M3}$,
Eqs. \ref{eq:rho_single_exp} and \ref{eq:hz_flare}), while the radial
surface density is given by a broken exponential:
\begin{equation}
  \label{eq:broken_exp_sigma}
  \Sigma(R) = \left\{
    \begin{array}{ll}
      \frac{1}{A_1}e^{-R/h_{R,\mathrm{in}}}, & \textrm{if}\ R\leq R_{\mathrm{peak}}\\
      \frac{1}{A_2}e^{-R/h_{R,\mathrm{out}}}, & \textrm{if}\ R> R_{\mathrm{peak}},
    \end{array} \right.
\end{equation}
where $h_{R,\mathrm{in}}$, $h_{R,\mathrm{out}}$ and
$R_{\mathrm{peak}}$ are free parameters, while $A_1$ and $A_2$ are
determined by the normalization condition, Eq.~(\ref{eq:normal}), and
equating the two components at $R=R_{\mathrm{peak}}$. In order to
allow the inner component to be an increasing function of radius, the
inner slope $h_{R,\mathrm{in}}$ is allowed to assume positive or
negative values. These models are listed in Table \ref{tab:models}.

\begin{table*}
  \centering
  \begin{tabular}{|c|c|c|}
    \hline
    $\nu_\star(R,z|\theta) = \Sigma(R|\theta)\zeta(z|R,\theta) $ & $\zeta(z|R,\theta)$
    & $\Sigma(R|\theta)$ \\
    \hline
    \hline
    {\bf M1} & double exp., Eq. (\ref{eq:rho_double_exp}) &
                                                            \multirow{4}{*}{single exp., Eq.~(\ref{eq:Sigma_single_exp})}\\
    {\bf M2} & single exp., Eq.~(\ref{eq:rho_single_exp}) & \\
    {\bf M3} & single flaring exp., Eqs.~(\ref{eq:rho_single_exp}) and (\ref{eq:hz_flare}) & \\
    {\bf M4} & double fl. inner exp., Eqs.~(\ref{eq:rho_double_exp}) and (\ref{eq:hz_flare})  & \\
    \hline
    {\bf M5} & single flaring exp., Eqs.~(\ref{eq:rho_single_exp}) and (\ref{eq:hz_flare}) & broken exp., Eq.~(\ref{eq:broken_exp_sigma})\\
    \hline \\
  \end{tabular}
  \caption{Summary of models fitted to the simulation data.}
  \label{tab:models}
  \end{table*}

  In order to test these models, in this work we perform Bayesian
  analyses where $\nu_\star(R,z|\theta)$ represents the probability to
  find a star particle at radius $R$ and absolute latitude $|z|$ or,
  considering the data altogether, the probability distribution for
  the data to be a sample realization of the model, given a fixed set
  of parameters $\theta$. In this approach, the likelihood function is
  defined as ${\cal L}\equiv \Pi_i \nu_{\star}(R_i,z_i|\theta)$ and,
  more conveniently, we define the log-likelihood
\begin{equation}
  \label{eq:lnL}
  \ln {\cal L}(\theta) = \sum_i\ln \left[\nu_{\star}(R_i,z_i|\theta)\right],
\end{equation}
where the sum runs over all star particles.

For all models described above, the log-likelihood is first maximized
with a downhill simplex algorithm \citep[][]{10.1093/comjnl/7.4.308}
in order to obtain an initial estimate of the best fit
parameters. These parameter values are then used as starting points to
MCMC-sample the posterior distribution function with the {\sc emcee}
package \citep[][]{2013PASP..125..306F}. The MCMC-sampling allows us
to better explore the parameter space and to estimate the
uncertainties on the best fit parameters. In all analyses in this
paper, the reported values of best fit parameters and their
uncertainties are estimated as the median and the interval containing
$68\%$ of the parameter sample, respectively.

In cases where more than one model is tested, the model comparison is
made by means of the so-called Bayes factor
\begin{equation}
  \label{eq:bayes_factor}
  \ln B=\ln \left(\frac{E_{1}}{E_{2}}\right),
\end{equation}
with positive (negative) values favouring model 1 (2). In this
definition, $E_1$ and $E_2$ are the statistical evidences of each
model, which in general can be approximated by
\begin{equation}
  \label{eq:evidence}
  \ln E \approx \ln {\cal L}_{bf} - \frac{k}{2}\ln N,
\end{equation}
where ${\cal L}_{bf}$ is the likelihood function evaluated in the best
fit model, $N$ is the number of data points (star particles) and $k$
is the number of parameters in the model (the test penalizes models
with larger numbers of parameters) -- see
e.g. \cite{2014sdmm.book.....I}.

According to \cite{doi:10.1080/01621459.1995.10476572}, values
$0 \lesssim \ln B_{12} \lesssim 1$ deserve only a bare mention in
favour of model 1, but model 1 is: weakly favoured for
$1 \lesssim \ln B_{12} \lesssim 3$, strongly favoured for
$3 \lesssim \ln B_{12} \lesssim 5$ and very strongly favoured for
$\ln B_{12} \gtrsim5$ (and the same for negative values in favour of
model 2).
  
\section{Results}
\label{sec:results}
\subsection{Preliminaries: the geometric thick disc}
\label{sec:all_sample_thick_disc}
In this section, we preliminarily show that the vertical density
profile generated by the simulation has, broadly speaking, all the
observed features of that of the Milky Way.

\begin{figure*}
  \includegraphics[width=\textwidth]{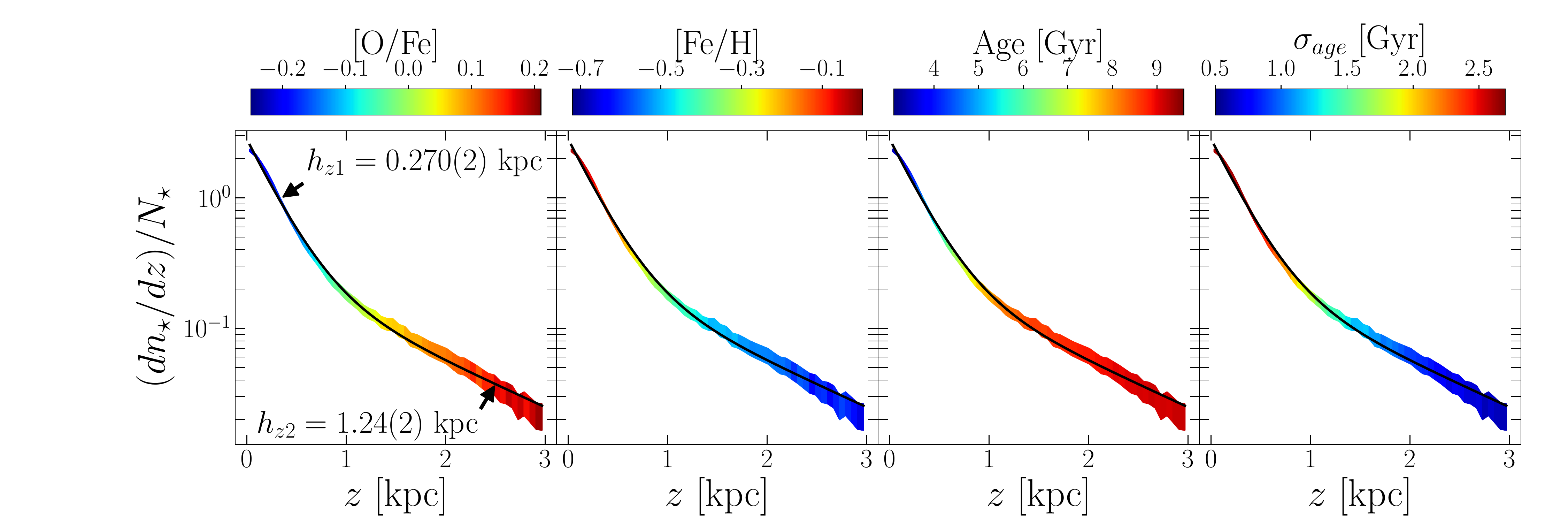}
  \caption{Density profile of the complete sample of star particles in
    the Solar neighborhood $6<R<10$ kpc, $|z|<3$ kpc. From left to
    right, the plots are colour-coded by [O/Fe], [Fe/H], age and age
    dispersion, respectively. Similarly to the MW
    \citep[see][]{2008ApJ...673..864J}, the data are well described by
    a double exponential model defining the geometric thin and thick
    discs, with scale height values shown in the left panel. The thick
    component ($|z|\gtrsim 1$kpc) is characterized by $\alpha$-rich,
    metal-poor and old populations, while the thin component
    ($|z|\lesssim 1$kpc) is $\alpha$-poor, metal-rich, young on
    average but with large age dispersions.}
  \label{fig:FB10_total_rho_z}
\end{figure*}

We select star particles at $t=10\ \mathrm{Gyr}$ within the cylinder
${6<R<10}$ kpc, ${|z|<3}$ kpc, resulting in a sample with
${N_\star = 368,021}$ particles. Fig.~\ref{fig:FB10_total_rho_z} shows
the normalized vertical density profile $(dn_\star/dz)/N_\star$ (where
$n_\star$ is the number of star particles in each bin) in log-scale,
split in 100 bins, with contours showing the Poisson uncertainty
multiplied by 5. These plots are colour-coded by mean [O/Fe], mean
[Fe/H], mean age and age dispersion, from left to right.

The vertical density profile has the characteristic double exponential
shape observed in the MW \citep[][]{1983MNRAS.202.1025G}, defining the
geometric thin and thick discs. As in the MW, the geometric thin disc
($|z|\lesssim 1\ \mathrm{kpc}$) is dominated by stars that are poor in
$\alpha$ elements (traced in the simulation by [O/Fe]), metal-rich and
young. On the other hand, the geometric thick disc
($|z|\gtrsim 1\ \mathrm{kpc}$) is characterized by $\alpha$-rich,
metal-poor and old stars. The right panel shows that while the thin
disc is composed of stars with significantly different ages, the thick
disc is much more uniform in age.

The (non-binned) data is then fit by model ${\bf M1}$,
Eqs.~(\ref{eq:Sigma_single_exp}) and (\ref{eq:rho_double_exp}). We
assume flat priors for the inverse of parameters $h_{z1}$ and $h_{z2}$
and flat priors for $\beta$. The best fit values of the two vertical
scale heights are $h_{z1}=0.270 \pm 0.002$ kpc and
$h_{z2}=1.24\pm 0.02$ kpc, as indicated in the left panel of
Fig.~\ref{fig:FB10_total_rho_z}. For the Milky Way,
\cite{2008ApJ...673..864J} found $h_{z1}\approx 0.3$ kpc and
$h_{z2}\approx 0.9$ kpc.

\begin{figure*}
  \includegraphics[width=\textwidth]{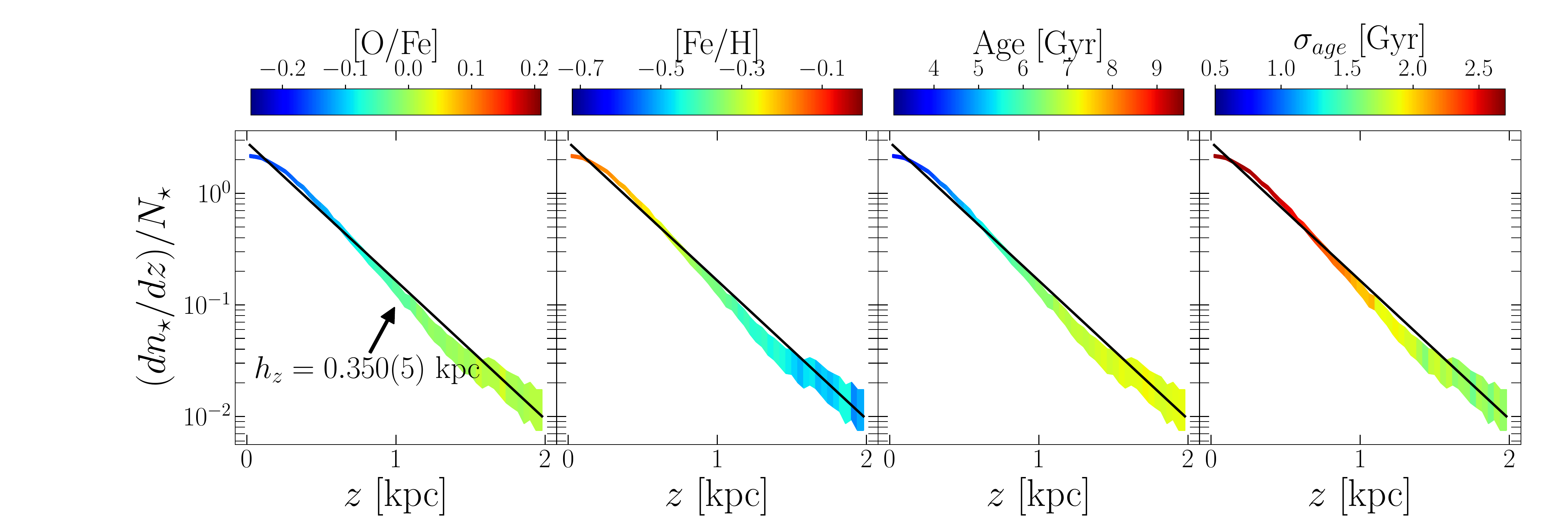}
  \caption{Density profile of the complete sample of star particles in
    the Solar neighborhood $6<R<10$ kpc, $|z|<3$ kpc for the
    simulation {\bf FB80}. From left to right, the plots are
    colour-coded by [O/Fe], [Fe/H], age and age dispersion,
    respectively. We see that the data is described by a single
    exponential (black line), i.e., different from the MW, with no
    geometric thick disc.}
  \label{fig:FB80_total_rho_z}
\end{figure*}

For comparison, in Fig. \ref{fig:FB80_total_rho_z} we show the
vertical density profile generated in the simulation in which the
stellar feedback efficiency is set to $80 \%$. The plots are
color-coded in the same way as in Fig. \ref{fig:FB10_total_rho_z}. In
this case, the vertical density profile can be fitted with a single
exponential with scale height $h_z \approx 0.3$ kpc, compatible with
the thin disc of the simulation {\bf FB10}. In other words, the
geometric thick disc is not formed in the non-clumpy simulation {\bf
  FB80}. For comparison of the subsequent results with those obtained
with simulation {\bf FB80}, see the Appendix.

\subsection{Density profiles of mono-abundance populations}
\label{sec:maps}
In this section, we start analyzing the simulation data in more detail
to compare our results with observational results on the MW. In
particular, in order to compare our results with those of
\cite{2012ApJ...753..148B}, we apply geometrical and chemical cuts
similar to those applied by these authors. We begin selecting star
particles within the region $5<R<12$ kpc, $0.3<|z|<3$ kpc, and with
abundances -1.5 < [Fe/H] < 0.6 and -0.4 < [O/Fe] < 0.5, resulting in a
sample with $N_\star = 252,393$ particles. Then we split this sample
in bins of width $0.1$ dex in [Fe/H] and $0.05$ dex in [O/Fe], each
bin defining a mono-abundance population (${\bf MAP}$). We require a
minimum of $N=250$ particles in each bin for fitting the model, which
is done with model ${\bf M2}$, Eqs.~(\ref{eq:Sigma_single_exp}) and
(\ref{eq:rho_single_exp}).

\begin{figure*}
  \includegraphics[width=\textwidth]{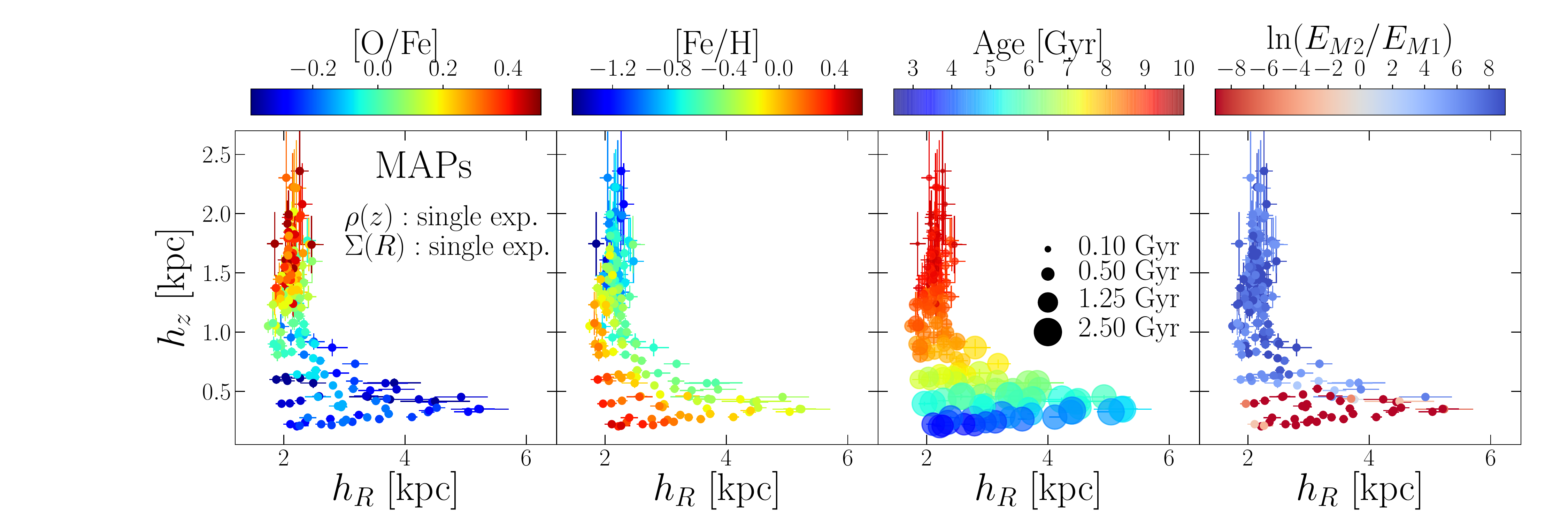}
  \caption{Best fit values of radial scale length $h_R$ vs vertical
    scale height $h_z$ for mono-abundance populations of star
    particles in $5<R<12$ kpc, $0.3<|z|<3$ kpc. From left to right,
    panels are colour-coded by [O/Fe], [Fe/H], mean age (point sizes
    meaning age dispersion) and by the Bayes factor comparing the fit
    quality of single vs double exponentials for the vertical density
    profile --see Eqs.~(\ref{eq:bayes_factor})-(\ref{eq:evidence}). In
    agreement with \protect\cite{2012ApJ...753..148B}, $\alpha$-rich
    and metal-poor MAPs are thick and centrally concentrated, as
    opposed to $\alpha$-poor and metal-rich MAPs, which are thin and
    spread out in radius. MAPs with older stars are more uniform in
    age and have vertical profiles better fitted by single
    exponentials, while younger MAPs have broader age distributions
    and are better fitted by double exponentials.}
  \label{fig:FB10_MAPs_single_exp_hR_hz}
\end{figure*}

Fig.~\ref{fig:FB10_MAPs_single_exp_hR_hz} shows the best fit values of
the scale length $h_R$ and scale height $h_z$ obtained with model
${\bf M2}$, colour-coded in various ways in the different panels. The
first noticeable feature in these plots is the strong anti-correlation
between $h_R$ and $h_z$: the thick populations (high $h_z$) are more
centrally concentrated (smaller $h_R$), while the thinner populations
are broadly distributed in radius (large $h_R$). The first and second
panels are colour-coded by [O/Fe] and [Fe/H], respectively. The thick
populations are dominated by stars that are $\alpha$-rich and
metal-poor, with a continuous decrease of [O/Fe] and increase of
[Fe/H] for the thinner populations. There is also a hint of a radial
metallicity gradient, but this is not perfect, with some metal-rich
populations being characterized by small $h_R$. These results are very
similar to those found by \cite{2012ApJ...753..148B} for the Milky
Way. See Fig. \ref{fig:FB80_MAPs_single_exp_hR_hz} and the Appendix
for a comparison with the non-clumpy simulation {\bf FB80}.

The third panel is colour-coded by age, with point sizes representing
the age dispersions in each MAP (black points show the scale). The
thick and centrally concentrated MAPs are characterized by old stars
and are very uniform in age. On the other hand, the thin populations
are young on average, but with a much broader age distribution, with
age dispersions as large as $2.5$ Gyr.

For a model comparison, we also fit model ${\bf M1}$, i.e. a double
exponential for the vertical density profile. The right panel of
Fig.~\ref{fig:FB10_MAPs_single_exp_hR_hz} is colour-coded by the Bayes
factor, Eqs.~(\ref{eq:bayes_factor})-~(\ref{eq:evidence}), comparing
models ${\bf M1}$ and ${\bf M2}$ (note that the best fit parameter
values shown are still the same as in the other panels, i.e. those of
model ${\bf M2}$). While most of the parameter space strongly favours
the single exponential (blue points), for some thin disc MAPs the
vertical density profile is better described by a double exponential
(red points). Comparison with the third panel shows that these are the
MAPs with larger age dispersions. Similarly to
\cite{2017ApJ...834...27M}, we conclude that in the same manner as the
vertical density profile of the complete sample,
Fig.~\ref{fig:FB10_total_rho_z}, seems to be characterized by a double
exponential as a consequence of superposing different populations, the
double exponentials required for the younger MAPs are also the
consequence of mixing different populations (with very different
ages).

\subsection{Density profiles of mono-age populations}
We now analyze the density profiles of co-eval (mono-age) populations
defined with 50 bins equally spaced in the logarithm of formation time
from 0.1 to 10 Gyr, and selected in a similar region as before
($5<R<12$ kpc), but now including the Galactic plane, i.e. $|z|<3$
kpc, in order to match the selection of \cite{2017MNRAS.471.3057M}. We
first focus on the vertical density profile of these populations for
which, in addition to models ${\bf M1}$ and ${\bf M2}$,
Eqs.~(\ref{eq:rho_double_exp}) and (\ref{eq:rho_single_exp}), we also
fit models ${\bf M3}$ and ${\bf M4}$, i.e. a single flaring
exponential and a double exponential allowing the inner component to
flare, respectively -- see Eq.~(\ref{eq:hz_flare}) and Table
\ref{tab:models}.

\begin{figure*}
  \includegraphics[width=\textwidth]{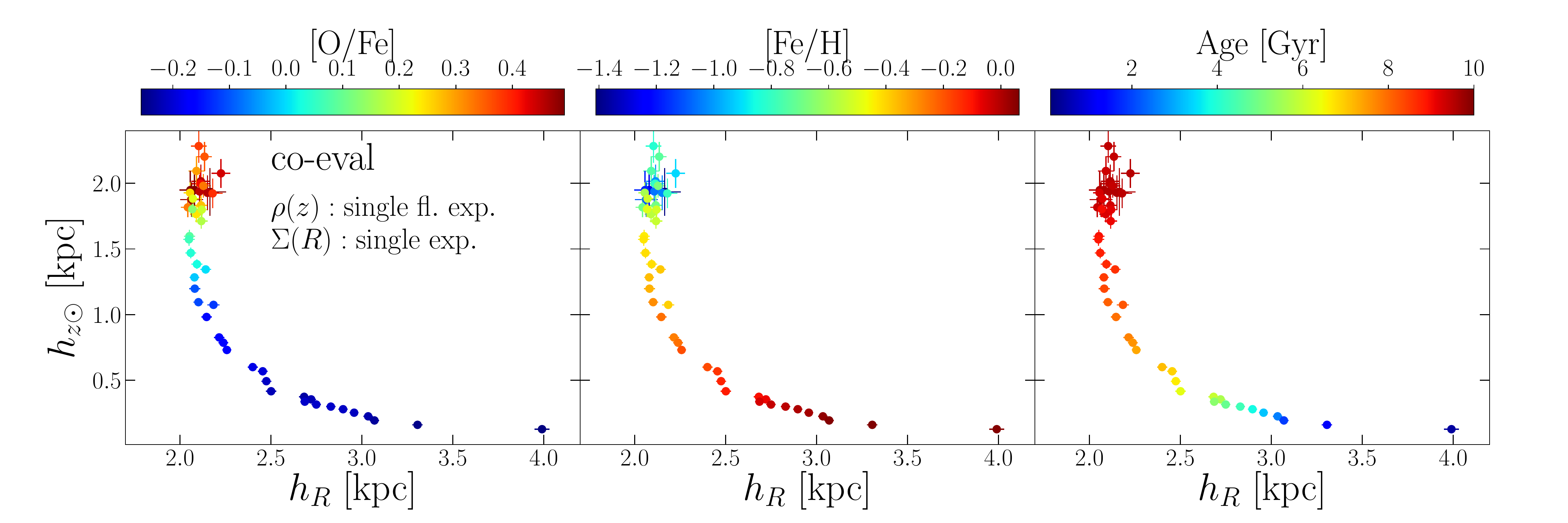}
  \caption{Best fit values of radial scale length $h_R$ vs vertical
    scale height at the Solar position $h_{z\odot}$ for mono-age
    populations obtained fitting a single flaring exponential
    model. Star particles are selected in $5<R<12$ kpc, $|z|<3$
    kpc. The anti-correlation between $h_R$ and $h_{z\odot}$ and the
    trends with [O/Fe], [Fe/H] and age are similar to, but cleaner
    than, those for MAPs --
    Fig.~\ref{fig:FB10_MAPs_single_exp_hR_hz}.}
  \label{fig:FB10_coeval_single_exp_hR_hz_I}
\end{figure*}

\begin{figure*}
  \includegraphics[width=\textwidth]{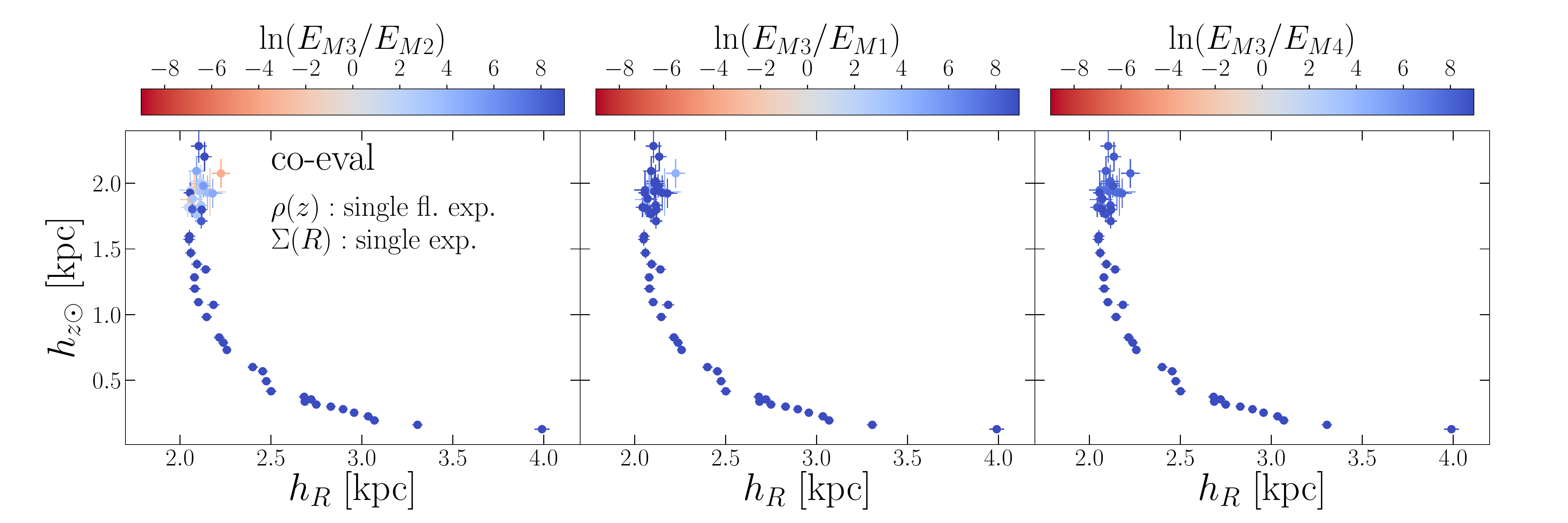}
  \caption{Same as Fig.~\ref{fig:FB10_coeval_single_exp_hR_hz_I}, but
    now colour-coded by the Bayes factor comparing the fit quality of
    the flaring single exponential (model {\bf M3}) with (from left to
    right) a non-flaring exponential (model {\bf M2}), a non-flaring
    double exponential ({\bf M1}) and a double exponential flaring the
    thin component ({\bf M4}). The flaring single exponential is
    favoured (blue points) in almost all mono-age populations, except
    for a few thick, old, $\alpha$-rich and metal-poor populations
    (see Fig.~\ref{fig:FB10_coeval_single_exp_hR_hz_I}), which show
    weak evidence in favour of a non-flaring single exponential (grey
    and red points in the left panel).}
  \label{fig:FB10_coeval_single_exp_hR_hz_II}
\end{figure*}

Fig~\ref{fig:FB10_coeval_single_exp_hR_hz_I} shows the best fit values
of parameters $h_R$ and $h_{z\odot}$ obtained with model ${\bf
  M3}$. The two left panels are again colour-coded by [O/Fe] and
[Fe/H] and the third panel is again colour-coded by age. We note an
anti-correlation between $h_R$ and $h_{z\odot}$ which is similar to,
but much clearer than, that of the MAPs in
Fig.~\ref{fig:FB10_MAPs_single_exp_hR_hz}. The thicker (high
$h_{z\odot}$) populations are again old, $\alpha$-rich and
metal-poor. The metallicity radial gradient is also much more clear
than that of the MAPs (middle panel). This clean anti-correlation
between scale length and scale height, with the respective trends with
abundances and ages, is similar to those found by
\cite{2013MNRAS.436..625S} using cosmological
simulations. Fig. \ref{fig:FB80_coeval_single_exp_hR_hz_I} in the
Appendix shows the equivalent results for the non-clumpy simulation
{\bf FB80}, where we conclude that, besides producing the geometric
thick disc by converting in-plane to vertical motion, the clumps are
also important to increase the radial scale length $h_R$ of old,
$\alpha$-rich populations due to in-plane scattering -- see the
Appendix.

The panels in Fig.~\ref{fig:FB10_coeval_single_exp_hR_hz_II} show the
same parameter values as
Fig.~\ref{fig:FB10_coeval_single_exp_hR_hz_I}, but now colour-coded by
the Bayes factor, Eqs.~(\ref{eq:bayes_factor})-(\ref{eq:evidence}),
comparing the quality of fits of model ${\bf M3}$ (single flaring
exponential for the vertical profile) with model ${\bf M2}$ (single
non-flaring exponential, left), model ${\bf M1}$ (double non-flaring
exponential, middle) and model ${\bf M4}$ (double inner-flaring
exponential, right). Model ${\bf M3}$ is strongly favoured with
respect to the other models in almost all the co-eval populations
(blue points), with a few exceptions for some of the thick populations
(high $h_{z\odot}$), which show either a weak evidence in favour of
this model or a weak evidence in favour of model ${\bf M2}$ (grey and
red points in the left panel).

Having shown that the discs generated by our simulation, and selected
by age, generally flare, i.e. get thicker for larger radii, with the
vertical density profiles being generally described by single
exponentials, we now shift our attention to the radial surface density
profiles that, up to this point, were modeled as single exponentials,
Eq.~(\ref{eq:Sigma_single_exp}). An important step in the study of the
density profiles of the MW was done by \cite{2016ApJ...823...30B} who,
analyzing red-clump stars from the APOGEE survey, identified that high
[$\alpha$/Fe] MAPs have radial surface density profiles well described
by single exponentials, while low [$\alpha$/Fe] MAPs show a break in
their profiles, with the inner part increasing with radius. Breaks in
the radial density profile have been observed also by e.g.
\cite{2006A&A...454..759P} in a sample of nearby galaxies from the
Sloan Digital Sky Survey (SDSS), by \cite{2007ApJ...667L..49D} in the
edge-on galaxy NGC 4244 and by \cite{2008ApJ...675L..65R} in
hydrodynamical simulations of self-consistent isolated galaxies -- see
\cite{2017ASSL..434...77D}. Similar features were later found for both
MAPs and mono-age populations in observational and simulation-based
results \citep[see e.g.][]{2015ApJ...804L...9M, 2017MNRAS.471.3057M}.

\begin{figure*}
  \includegraphics[width=\textwidth]{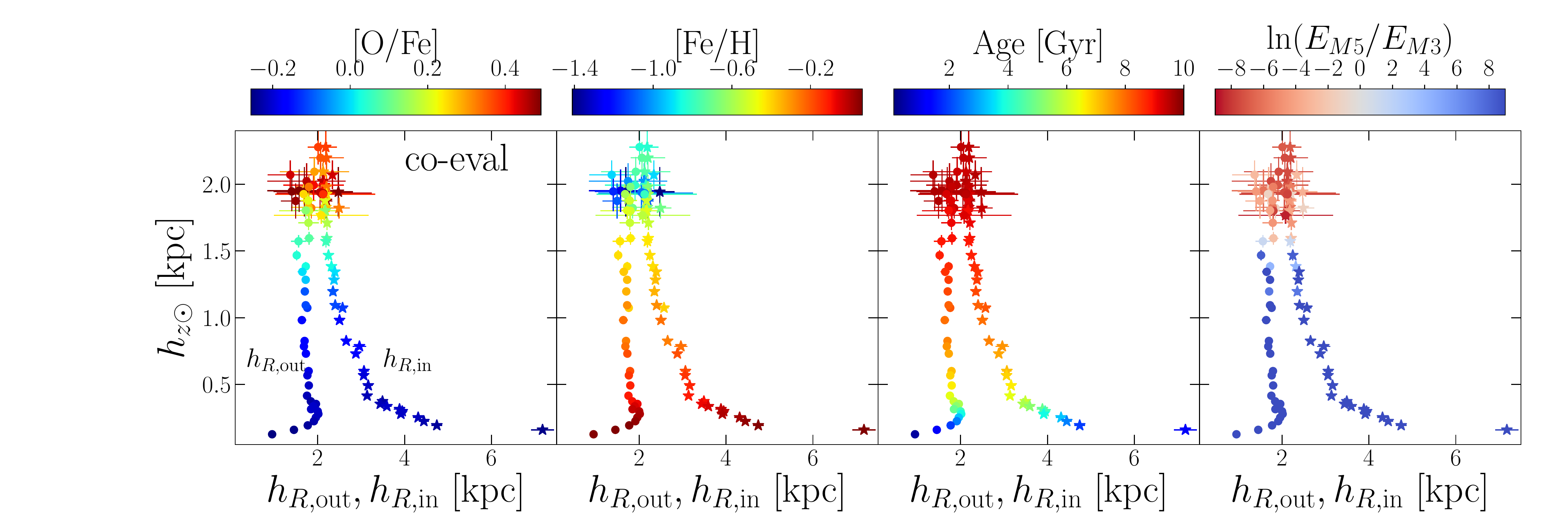}
  \caption{Best fit values of the outer and inner radial scale lengths
    $h_{R,\mathrm{out}}$, $h_{R,\mathrm{in}}$ vs vertical scale height
    at the Solar position $h_{z\odot}$ for mono-age populations
    obtained fitting a single flaring exponential model for
    $\zeta(z|R)$ and a broken exponential for $\Sigma(R)$ -- model {\bf
      M5}. The data selection, model fit and colour-coding are the
    same as that of
    Fig.~\ref{fig:FB10_coeval_single_exp_hR_hz_I}. Young,
    $\alpha$-poor and metal-rich populations have sharp breaks,
    i.e. significant differences
    $h_{R,\mathrm{in}} - h_{R,\mathrm{out}}$, while old,
    $\alpha$-rich, metal-poor populations have
    $h_{R,\mathrm{out}} \approx h_{R,\mathrm{in}}$, and thus flat
    exponential profiles. The right panel is colour-coded by the Bayes
    factor comparing the single versus broken exponential models for
    $\Sigma(R)$. High-[O/Fe] populations are better described by
    single exponentials (red points), while low [O/Fe] populations
    strongly favour the broken exponential model (blue points).}
  \label{fig:FB10_coeval_broken_exp_hR_hz}
\end{figure*}

We test these features by means of model ${\bf M5}$,
Eq.~(\ref{eq:broken_exp_sigma}), which is characterized by a broken
exponential, i.e. an exponential with slope $h_{R,\mathrm{in}}$ for
$R\leq R_\mathrm{peak}$ and another exponential with slope
$h_{R,\mathrm{out}}$ for $R
>R_\mathrm{peak}$. Fig.~\ref{fig:FB10_coeval_broken_exp_hR_hz} shows
the best fit values of the parameters $h_{R,\mathrm{in}}$ and
$h_{R,\mathrm{out}}$ versus $h_{z\odot}$. The three left panels are
colour-coded as in the previous figures. Note that for the young,
$\alpha$-poor and metal-rich populations, the parameters
$h_{R,\mathrm{in}}$ and $h_{R,\mathrm{out}}$ assume substantially
different values, which already suggests that the data of these
populations really require a broken (instead of a single)
exponential. Essentially all populations assume positive values for
$h_{R,\mathrm{in}}$ (there is only one population for which we found
$h_{R,\mathrm{in}} \approx -70$ kpc, whose profile is basically flat
in the inner part), which means that, despite the break, their inner
profiles are still decreasing functions of radii. Except for the very
young, (age $\lesssim 2$ Gyr), all populations approximately have a
common $h_{R,\mathrm{out}} \approx 2$ kpc. On the other hand, the
parameter $h_{R,\mathrm{in}}$ is anti-correlated with $h_{z\odot}$,
similarly to that observed fitting a single exponential (see
Fig.~\ref{fig:FB10_coeval_single_exp_hR_hz_I}).

Going up in these panels, i.e. for higher $\alpha$ and age and lower
metallicities, the parameter values of $h_{R,\mathrm{in}}$ approach
those of $h_{R,\mathrm{out}}$, until they merge into the
high-uncertainty region of the $\alpha$-rich, old and metal-poor
populations. The right panel compares models ${\bf M5}$ and
${\bf M3}$, i.e. the broken versus single exponential for the surface
density $\Sigma(R)$, while keeping a single flaring exponential for
$\zeta(z|R)$, Eqs.~(\ref{eq:rho_single_exp}) and
~(\ref{eq:hz_flare}). In agreement with the initial suggestion, it is
clear that the thicker ($\alpha$-rich and old) populations are better
described by single exponentials (red points), while the thinner ones
($\alpha$-poor and younger) are better described by broken
exponentials (blue points). This also explains the large uncertainties
in $h_{R,\mathrm{in}}$, $h_{R,\mathrm{out}}$ and $h_{z\odot}$ in the
thick populations, for which the broken exponential model over-fits
the data. See Fig.~\ref{fig:FB80_coeval_broken_exp_hR_hz} for a
comparison with simulation {\bf FB80}.

\begin{figure}
  \includegraphics[width=\columnwidth]{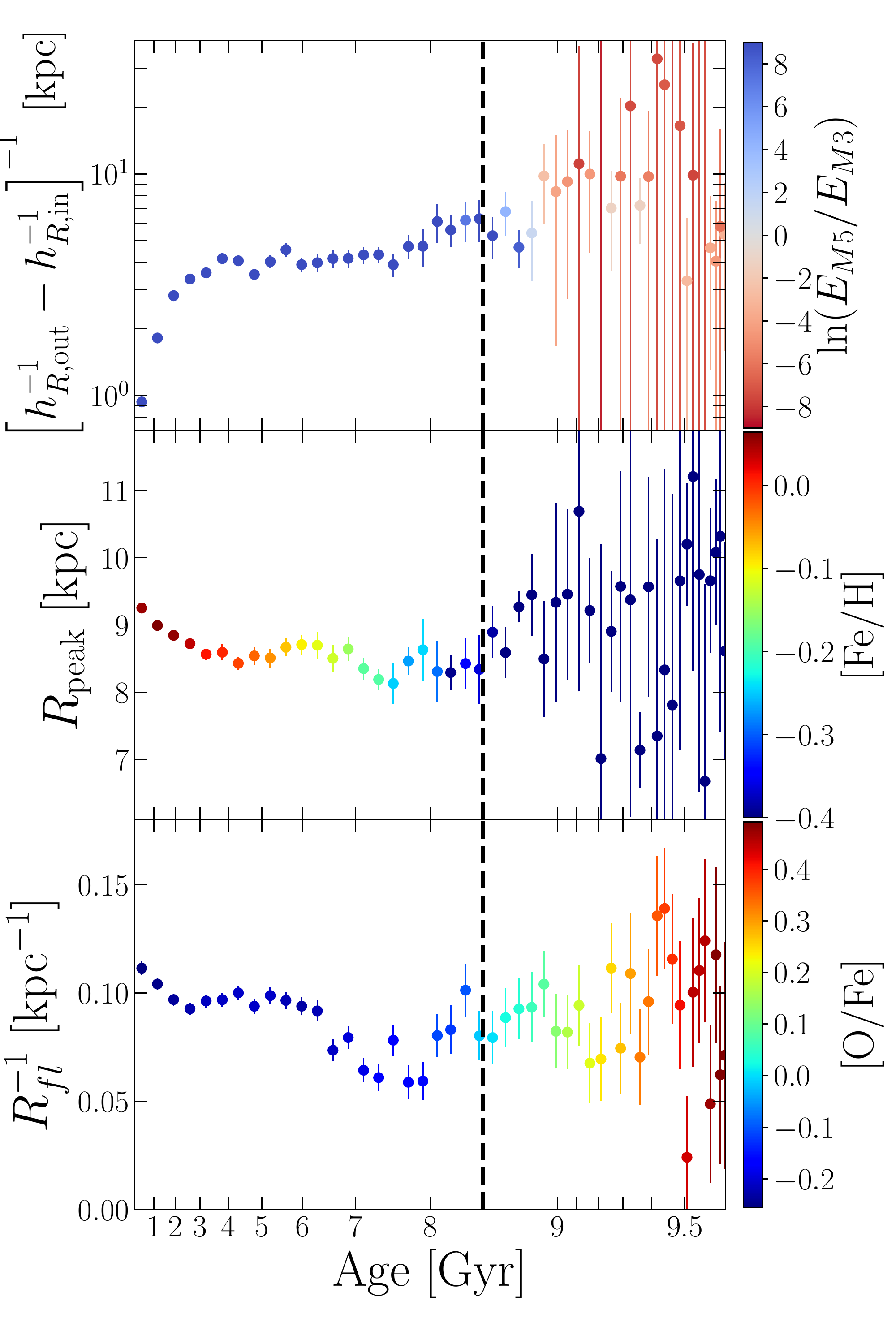}
  \caption{Top: the quantity $\Delta$, Eq.~(\ref{eq:delta}), proposed
    by \protect\cite{2017MNRAS.471.3057M} for measuring how sharply
    broken $\Sigma(R)$ is (a small value indicates a sharp peak),
    colour-coded by the Bayes factor comparing a single and a broken
    exponential models. Center: the position of the break in
    $\Sigma(R)$, $R_{\mathrm{peak}}$, colour-coded by [Fe/H]. Bottom:
    inverse flaring radius (a small value means weak flaring),
    colour-coded by [O/Fe]. The vertical dashed line crosses the
    bottom panel at [O/Fe]=0 (corresponding to an age $\approx 8.5$
    Gyr), defining $\alpha$-poor (left) and $\alpha$-rich (right)
    regions with very different behavior for all three parameters:
    low-$\alpha$ populations are described by a broken $\Sigma(R)$,
    with peak sharpness $\Delta$ decreasing with age, peak position
    decreasing with age for ages $\lesssim 4$ Gyr and presenting small
    wiggles for larger ages, and flaring level $R_{fl}^{-1}$ overall
    decreasing with age. High-$\alpha$ populations have $\Sigma(R)$
    better described by single exponentials, thus with large
    uncertainties for the over-fitting quantities $\Delta$ and
    $R_\mathrm{peak}$. Despite the large uncertainties, $R_{fl}^{-1}$
    seems to increase weakly with age for the high-$\alpha$ region.}
  \label{fig:FB10_coeval_broken_exp_Rpeak_Rfl}
\end{figure}

Fig.~\ref{fig:FB10_coeval_broken_exp_Rpeak_Rfl} shows the best fit
values of the other parameters characterizing the surface density
$\Sigma(R)$ as a function of age, obtained with model ${\bf M5}$. The
vertical line, approximately at $8.5$ Gyr, divides the samples in
low-$\alpha$ (left) and high-$\alpha$ (right) populations, with [O/Fe]
abundances lower and larger than zero, respectively. The top panel
shows, colour-coded by the Bayes factor, the quantity
\begin{equation}
  \label{eq:delta}
  \Delta = \left[h_{R,\mathrm{out}}^{-1} - h_{R,\mathrm{in}}^{-1}\right]^{-1},
\end{equation}
as proposed by \cite{2017MNRAS.471.3057M}, and intended to diagnose
how sharply broken $\Sigma(R)$ is. A low value of $\Delta$ means a
sharp break, while large values mean more broad and continuous
profiles (for a single exponential, that quantity tends to
infinity). We see that the youngest populations ($\lesssim 2$ Gyr)
have sharp breaks, but the profiles get smoother very quickly after
$2$ or $3$ Gyr of evolution, with a more gentle smoothing in the later
evolution. However, despite this smoothing, a noticeable break is
still present for populations as old as $8$ or $9$ Gyr, as indicated
by the Bayes factor favouring model ${\bf M5}$ (blue points). This
smooth trend is dramatically different for populations older than $9$
Gyr which are better described by single exponentials (red points).

\begin{figure}
  \center
  \includegraphics[scale=0.35]{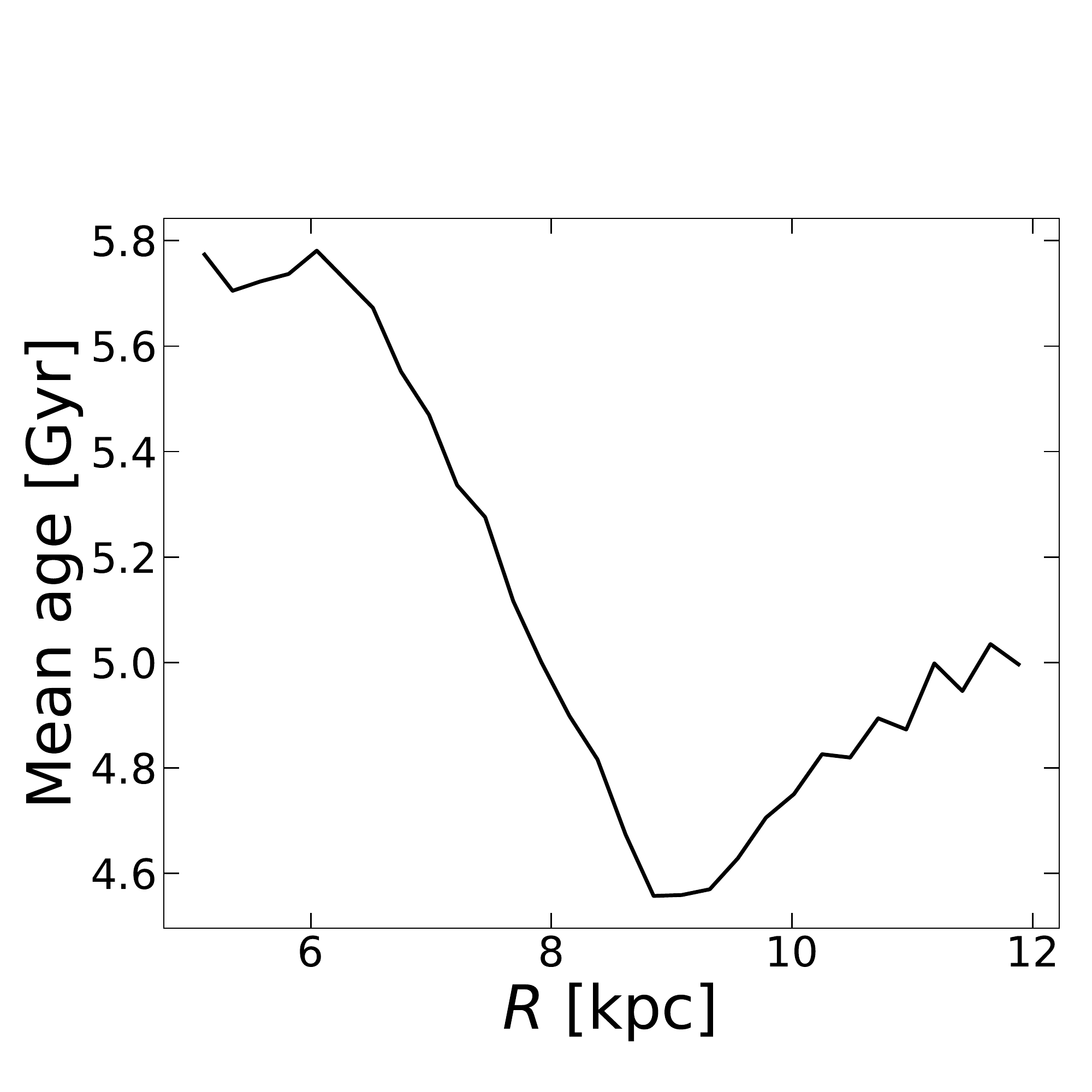}
  \caption{The mean stellar age at each radius obtained with the
    simulation {\bf FB10}. As found by
    \protect\cite{2008ApJ...675L..65R}, the position of the minimum,
    here at $R\approx 9$ kpc, correlates well with the position of the
    break in $\Sigma(R)$ for the youngest population, shown in the
    central panel of Fig. \ref{fig:FB10_coeval_broken_exp_Rpeak_Rfl}.}
  \label{fig:FB10_mean_age_vs_R}
\end{figure}

The central panel of Fig.~\ref{fig:FB10_coeval_broken_exp_Rpeak_Rfl}
shows the position of the break, $R_\mathrm{peak}$, colour-coded by
[Fe/H]. $R_\mathrm{peak}$ does not depend strongly on age, assuming
values between $8$ and $9$ kpc for the $\alpha$-poor populations. For
the youngest (age $\lesssim 3$ Gyr) and more metal-rich populations,
$R_\mathrm{peak}$ decreases for decreasing metallicities. For larger
ages, and lower metallicities, $R_\mathrm{peak}$ seems to have some
wiggles. Since the $\alpha$-rich populations are better described by
single exponentials, the uncertainties on $R_\mathrm{peak}$ are large,
but there is a hint for an increasing $R_\mathrm{peak}$ with
decreasing metallicities. In their self-consistent hydro-dynamical
simulation, \cite{2008ApJ...675L..65R} found a tight correlation
between the position of the break and the minimum of the mean age (as
a function of radii $R$). In Fig.~\ref{fig:FB10_mean_age_vs_R}, which
shows the mean age as a function of $R$ in our simulation, we observe
a minimum at $R \approx 9$ kpc, which is similar to the values of
$R_\mathrm{peak}$ obtained for the youngest co-eval populations shown
in the central panel of
Fig.~\ref{fig:FB10_coeval_broken_exp_Rpeak_Rfl}.

The bottom panel of Fig. \ref{fig:FB10_coeval_broken_exp_Rpeak_Rfl}
shows the inverse of the typical flaring scale $R_{fl}^{-1}$
colour-coded by [O/Fe]. Small values of $R_{fl}^{-1}$ indicate the
absence of a flare, and large values indicate significant
flaring. Even for the low-$\alpha$ populations, the dependence on age
is not trivial, but we note that the overall $R_{fl}^{-1}$ decreases
with age for these populations. For the high-$\alpha$ populations,
compatible with the fact that for these populations we have weak or no
statistical significance in support of the flaring model (see left
panel of Fig.~\ref{fig:FB10_coeval_single_exp_hR_hz_II}), the
uncertainties are large, but overall $R_{fl}^{-1}$ seems to increase
with age. Fig.~\ref{fig:FB80_coeval_broken_exp_Rpeak_Rfl} in the
Appendix shows the equivalent plots for {\bf FB80}.

\section{Comparison with previous results}
\label{sec:compare_results}
We show in Sec. \ref{sec:all_sample_thick_disc} that the discs
generated by the simulation have gross properties in reasonable
agreement with data from the MW \citep[][]{2008ApJ...673..864J}, and
also the expected correlations with age and chemical abundances.

In Sec. \ref{sec:maps}, we showed that when selecting mono-abundance
populations and applying geometrical cuts similar to those used by
\cite{2012ApJ...753..148B}, we arrive at similar results when fitting
single exponentials for both the vertical and radial density profiles:
the vertical profile of most of the chemical space is well described
by single exponentials, as found by \cite{2012ApJ...753..148B} -- see
Fig.~\ref{fig:FB10_MAPs_single_exp_hR_hz}. The old, $\alpha$-rich and
metal poor MAPs are thick and more centrally concentrated, as opposed
to the MAPs which are young, $\alpha$-poor and metal-rich, which are
thin and radially spread out. The MAPs which are the youngest on
average have larger age dispersions and are better described by double
exponentials.

Using cosmological simulations and selecting co-eval (mono-age)
populations, \cite{2013MNRAS.436..625S} found an anti-correlation
between $h_R$ and $h_z$, and trends with chemical abundances and ages
which are similar to, but much cleaner than, those of MAPs. With our
isolated galaxy simulation we arrive at very similar results -- see
Fig.~\ref{fig:FB10_coeval_single_exp_hR_hz_I}.

In a detailed study of nearby galaxies, \cite{2011ApJ...741...28C}
found no evidence for significant flaring of thick
discs. \cite{2016ApJ...823...30B}, analyzing red-clump stars from the
APOGEE survey, found that the vertical density profiles of thick,
$\alpha$-rich MAPs in their sample were described by single
non-flaring exponentials, while the thin, $\alpha$-poor MAPs were
better described by single flaring
exponentials. \cite{2015ApJ...804L...9M}, selecting MAPs from
cosmological simulations also arrived at a similar conclusion. When
selecting co-eval populations, \cite{2015ApJ...804L...9M} found that
all populations are described by single flaring exponentials, with a
correlation between age and flaring, i.e. that young (old) populations
flare less (more).

More recently, \cite{2017MNRAS.471.3057M} using data from the APOGEE
survey, analyzed the spatial distribution of red giant stars in the
MW. They first split the data into $\alpha$-rich and $\alpha$-poor
groups and for each group they selected mono-age, mono-metallicity
populations. They also found that all populations are well described
by single exponentials with different flaring levels:
high-[$\alpha$/Fe] populations tend to flare less in general, with
flaring being detected with weaker statistical significance, while
low-[$\alpha$/Fe] show a clear, statistically significant
flaring. Additionally, \cite{2017MNRAS.471.3057M} found that the
$\alpha$-poor populations show an anti-correlation between age and
flaring, i.e. that younger (older) populations flare more (less),
while for the $\alpha$-rich they found a correlation, i.e. that
younger (old) populations flare less (more).

In our analysis, we conclude that the vertical density profiles of
old, high-$\alpha$ populations are well described by single
exponentials, with weak or no-evidence for flaring, while the
low-$\alpha$ populations are well described by single flaring
exponentials -- see Figs.~\ref{fig:FB10_coeval_single_exp_hR_hz_I} and
\ref{fig:FB10_coeval_single_exp_hR_hz_II}. The inverse of the flaring
radius decreases with age for $\alpha$-poor populations and increases
with ages (although with large uncertainties) for $\alpha$-rich
populations -- see the bottom panel of
Fig.~\ref{fig:FB10_coeval_broken_exp_Rpeak_Rfl}. All these results are
in good agreement with those of \cite{2017MNRAS.471.3057M} for the MW.

Regarding the radial surface density profiles, the $\alpha$-rich
co-eval populations are well described by single exponentials, while
the $\alpha$-poor populations are better described by broken
exponentials, also in agreement with \cite{2016ApJ...823...30B} and
\cite{2017MNRAS.471.3057M}. The profiles of younger populations are
more sharply broken, and get flatter with age.

A difference between our results and those of
\cite{2016ApJ...823...30B} and \cite{2017MNRAS.471.3057M} is that,
despite the breaks, we do not find any population with the inner
profile increasing with radius, i.e. with
$h_{R,\mathrm{in}}<0$. Another apparent difference is that
\cite{2017MNRAS.471.3057M} observe an anti-correlation between the
position of the break $R_{\mathrm{peak}}$ and the metallicity, while
in our case the trend is the opposite for the younger populations (age
$\lesssim 5$ Gyr). On the other hand, we find that overall
$R_{\mathrm{peak}}$ changes only by less than 1 kpc for the
low-[$\alpha$/Fe] populations and have large uncertainties for the
high-[$\alpha$/Fe] populations. However, note that our populations are
defined by ages only, i.e. we do not bin in [Fe/H] and thus averaging
could be the origin of the observed difference. Another possible
explanation for these differences is the fact that the simulated
galaxy does not produce a bar, whose contribution to shaping the
radial profile might be non-negligible.

Interestingly, running hydrodynamical simulations,
\cite{2008ApJ...675L..65R} found that the break of the radial density
profile occurs at the same radius for all stellar ages, i.e. that
$R_{\mathrm{peak}}$ does not depend on age (the same was observed by
\cite{2007ApJ...667L..49D} for NGC 4244). They also found that
$R_{\mathrm{peak}}$ coincides with the radius of minimum mean stellar
age and that on average
$R_{\mathrm{peak}}/h_{R,\mathrm{in}} \approx 2.6$, which is in good
agreement with the mean value $2.5$ found by
\cite{2006A&A...454..759P} in a sample of nearby galaxies observed by
SDSS and the value $2.3\pm 0.7$ found by \cite{2006ApJ...650..644E}
for high-redshift spiral galaxies. Selecting $\alpha$-poor stars (for
which the break is present), defined in
Fig.~\ref{fig:FB10_coeval_broken_exp_Rpeak_Rfl} as those stars younger
than $8.5$ Gyr, we obtain a mean value
$R_{\mathrm{peak}}/h_{R,\mathrm{in}} \approx 2.5$. As we can see, the
results obtained in this paper are in good agreement with the
observational and simulation-based results just mentioned.

Finally, in their (non-clumpy) simulation, \cite{2008ApJ...675L..65R}
concluded that the rising mean age in the external disc
($R>R_\mathrm{peak}$) is due to radial migration of relatively old
stars in nearly circular orbits (for which migration is more
efficient). A kinematical analysis of our simulation reveals that
approximately $\approx 85\%$ of star particles in the outer disc
($9 < R/\mathrm{kpc} < 12$) have eccentricity $< 0.4$, while
$\approx 5\%$ have eccentricity $>0.7$. Thus in our simulation we also
conclude that the outer disc is mainly composed of star particles in
nearly circular orbits (a significant fraction of which must have
migrated from the inner disc), with a small contribution of star
particles in highly eccentric orbits generated by early scattering in
the clumps.

\section{Conclusions}
\label{sec:conclusions}
In this paper, we analyze SPH simulations in order to investigate the
role of clumps for the evolution of the spatial structure of galactic
discs. A simulation with low feedback efficiency allows the formation
of clumps which scatter the old stars to high galactic latitudes and
sink to the center after $\approx 4$ Gyr. We show that this simulation
naturally gives rise to galactic discs (particularly a thick disc)
with properties in good agreement with those observed in the Milky
Way.

Selecting all stars in a given geometric cut, without any additional
cut in chemistry or age, the vertical density profile is characterized
by the typical double exponential law, with the same trends observed
in the Milky Way \citep[][]{2005A&A...433..185B, 2008ApJ...673..864J}:
the geometric thick disc is composed of stars that are old, rich in
[$\alpha$/Fe] and metal poor, while the thin disc is dominated by
stars that are young, [$\alpha$/Fe]-poor and metal-rich.

Applying chemical cuts, we observe that mono-abundance populations are
generally characterized by single exponential vertical density
profiles, with the exception of some young MAPs with large age
dispersions, which require a double exponential.

The selection of mono-age populations results in vertical density
profiles well described by single exponentials with very clean
correlations between the vertical scale height $h_z$ and radial scale
length $h_R$, as observed by \cite{2013MNRAS.436..625S}. The radial
density profiles of high-$\alpha$ mono-age populations are also single
exponentials, while those of low-$\alpha$ populations are broken
exponentials. The sharpness of the break decreases with age, which is
probably the consequence of radial migration erasing primeval
gradients, as pointed out by \cite{2017MNRAS.471.3057M}.

It is interesting to note that radial migration is also traditionally
believed to be responsible for the flaring of low [$\alpha$/Fe]
populations, i.e. for the disc thickening at larger radii for the
stars in the geometric thin disc, in which case one would expect young
populations to flare less. However, we find \citep[in agreement
with][]{2017MNRAS.471.3057M} the opposite trend, i.e. that the flaring
level, $R_{fl}^{-1}$, decreases with age for these populations -- see
Fig. \ref{fig:FB10_coeval_broken_exp_Rpeak_Rfl}. This fact is
interpreted by \cite{2017MNRAS.471.3057M} as suggestive that old
mergers have suppressed the flaring of the old populations in the
Milky Way. On the other hand, since we simulate an isolated galaxy,
our results show that internal dynamical processes are able to build
the observed trends, without the need for mergers.

The detailed agreement of the results presented in this work with
various observational results allows us to conclude that the
scattering of old, $\alpha$-rich and metal-poor stars by clumps formed
early in the Galaxy is a viable mechanism for the formation of the
thick disc.

\section*{Acknowledgements}
VPD and LBS are supported by STFC Consolidated grant \#~ST/R000786/1.
The simulations in this paper were run at the DiRAC Shared Memory
Processing system at the University of Cambridge, operated by the
COSMOS Project at the Department of Applied Mathematics and
Theoretical Physics on behalf of the STFC DiRAC HPC Facility
(www.dirac.ac.uk). This equipment was funded by BIS National
E-infrastructure capital grant ST/J005673/1, STFC capital grant
ST/H008586/1 and STFC DiRAC Operations grant ST/K00333X/1. DiRAC is
part of the National E-Infrastructure.



\bibliographystyle{mnras}
\bibliography{/Users/lbs/uclan/refs_lbs_uclan}

\appendix
\section{Comparison with a non-clumpy simulation}
In this section, we perform the same analysis as that of
Sec. \ref{sec:results}, but applied to the simulation {\bf FB80}, in
which the stellar feedback efficiency is set to $80\%$. As discussed
in Sec. \ref{sec:sims}, this high feedback efficiency inhibits the
formation of clumps and thus the comparison of results from
simulations {\bf FB10} and {\bf FB80} allows us to clarify the role of
the clumps. We already showed in Fig. \ref{fig:FB80_total_rho_z} that
the vertical density profile obtained with the simulation {\bf FB80}
is characterized by a single exponential, i.e. that this simulation
does not produce a geometric thick disc.

\begin{figure*}
  \includegraphics[width=\textwidth]{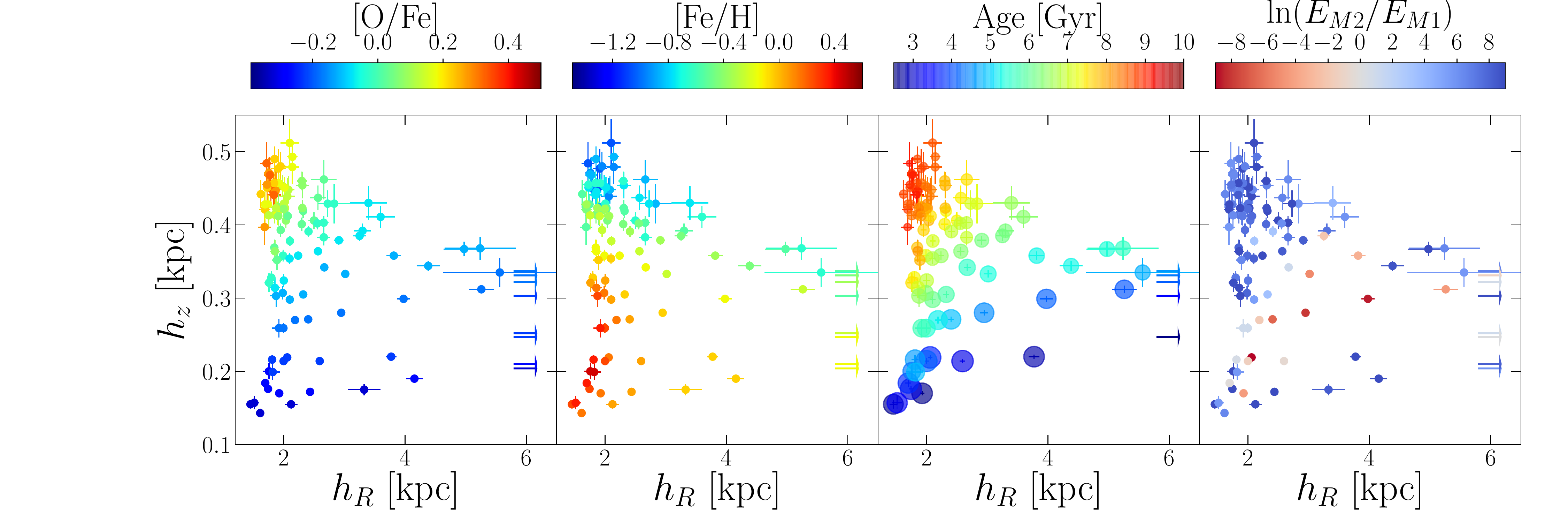}
  \caption{Best fit values of radial scale length $h_R$ vs vertical
    scale height $h_z$ for mono-abundance populations of star
    particles in $5<R<12$ kpc, $0.3<|z|<3$ kpc in the simulation {\bf
      FB80}. From left to right, panels are colour-coded by [O/Fe],
    [Fe/H], mean age (point sizes meaning age dispersion) and by the
    Bayes factor comparing the fit quality of single vs double
    exponentials for the vertical density profile --see
    Eqs.~(\ref{eq:bayes_factor})-(\ref{eq:evidence}). The
    anti-correlation between $h_R$ and $h_z$ observed in the
    simulation {\bf FB10} -- Fig.~\ref{fig:FB10_MAPs_single_exp_hR_hz}
    -- and in the MW \citep[][]{2012ApJ...753..148B} is blurred
    here. Note also that the vertical scale is smaller than that of
    Fig.~\ref{fig:FB10_MAPs_single_exp_hR_hz}.}
  \label{fig:FB80_MAPs_single_exp_hR_hz}
\end{figure*}

In Fig. \ref{fig:FB80_MAPs_single_exp_hR_hz}, we show the best fit
values of the scale length $h_R$ and scale height $h_z$ obtained
fitting single exponentials, Eqs. (\ref{eq:Sigma_single_exp}) and
(\ref{eq:rho_single_exp}), to mono-abundance populations (MAPs)
selected from the simulation {\bf FB80}. This figure is to be compared
to Fig.~\ref{fig:FB10_MAPs_single_exp_hR_hz}. First, we note that the
typical values of $h_z$ are significantly smaller than those in
Fig.~\ref{fig:FB10_MAPs_single_exp_hR_hz}, showing again the absence
of thicker components in {\bf FB80}. Additionally, the
anti-correlation between $h_z$ and $h_R$ observed for the simulation
{\bf FB10} and for the Milky Way \citep{2012ApJ...753..148B} is
blurred for {\bf FB80}, with the best fit parameter values being more
spread in the diagram.

\begin{figure*}
  \includegraphics[width=\textwidth]{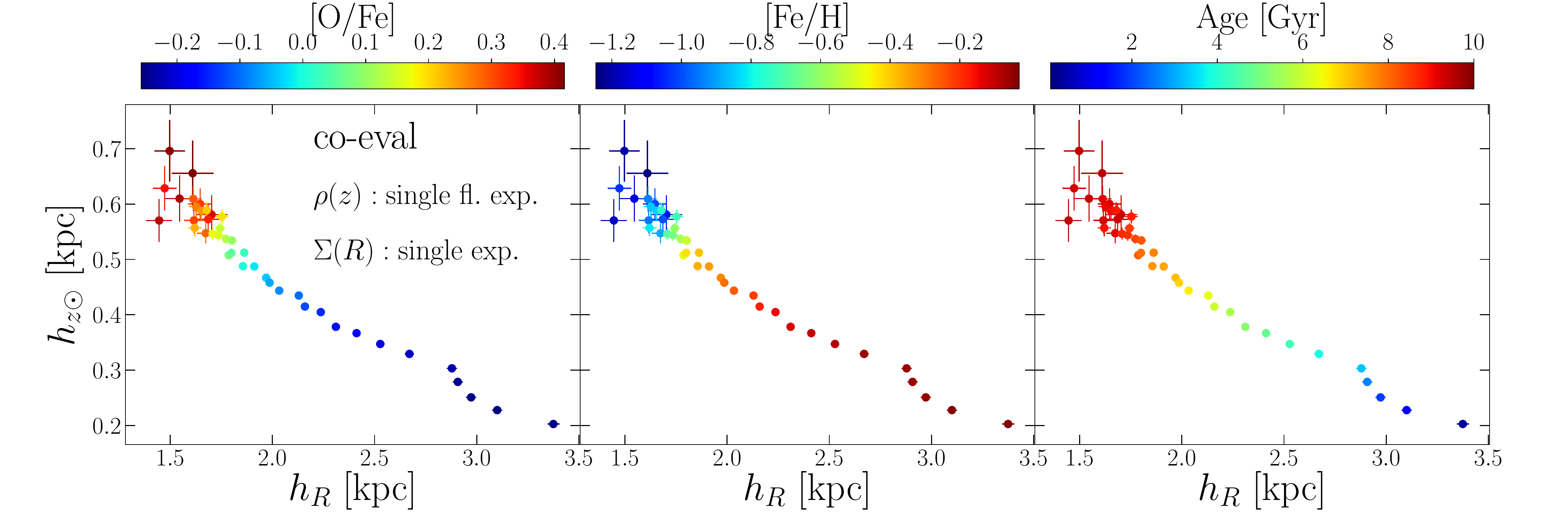}
  \caption{Best fit values of radial scale length $h_R$ vs vertical
    scale height at the Solar position $h_{z\odot}$ for mono-age
    populations obtained fitting a single flaring exponential model
    for the simulation {\bf FB80}. Star particles are selected in
    $5<R<12$ kpc, $|z|<3$ kpc. The anti-correlation between $h_R$ and
    $h_{z\odot}$ in this case is different from that observed in the
    simulation {\bf FB10} --
    Fig.~\ref{fig:FB10_coeval_single_exp_hR_hz_I} -- and by
    \protect\cite{2013MNRAS.436..625S} in cosmological simulations,
    with $\alpha$-rich populations assuming smaller values for both
    $h_{z\odot}$ and $h_R$.}
  \label{fig:FB80_coeval_single_exp_hR_hz_I}
\end{figure*}

Fig.~\ref{fig:FB80_coeval_single_exp_hR_hz_I} (to be compared to
Fig. \ref{fig:FB10_coeval_single_exp_hR_hz_I}), shows the best fit
values of the scale length $h_R$ and scale height at the Solar
position $h_{z\odot}$ obtained fitting the single flaring exponential
model, Eqs. (\ref{eq:Sigma_single_exp}), (\ref{eq:rho_single_exp}) and
(\ref{eq:hz_flare}), to the data of simulation {\bf FB80}. The main
difference with respect to
Fig. \ref{fig:FB10_coeval_single_exp_hR_hz_I} is again the typical
values of $h_{z\odot}$, which are significantly smaller for {\bf
  FB80}. It is interesting to note also that the thicker and older
components in {\bf FB80} assume $h_R$ values significantly smaller
than those in {\bf FB10}
(Fig. \ref{fig:FB10_coeval_single_exp_hR_hz_I}). A possible
interpretation for this is the following: in the clumpy simulation
{\bf FB10}, the old, $\alpha$-rich populations are scattered by the
clumps. In the reference frame of the clump, this scatter is
quasi-isotropic, i.e. scatters to any direction are equally
probable. Since the clumps are moving in the galactic plane, in the
reference frame of the galactic center this scattering should have
some anisotropy but, still, scatters that promote stars to high
latitudes $z$ also tend to increase the dispersion of radii $R$. In
this way, the absence of scattering in {\bf FB80}, apart from not
allowing the formation of thicker components, also does not allow the
promotion of the older stars, typically formed in inner radii
(inside-out growth) to larger radii. The direct consequence is that in
the non-clumpy simulation the scale lengths $h_R$ of old populations
are only affected by secular heating and are closer to their pristine
values and thus typically small. On the other hand, the youngest and
thinnest populations are not affected by the clumps, which have sunk
to the center by $\approx 4$ Gyr in {\bf FB10}. In this way, the
typical values of $h_R$ and $h_z$ for these populations are similar in
the two simulations.

\begin{figure*}
  \includegraphics[width=\textwidth]{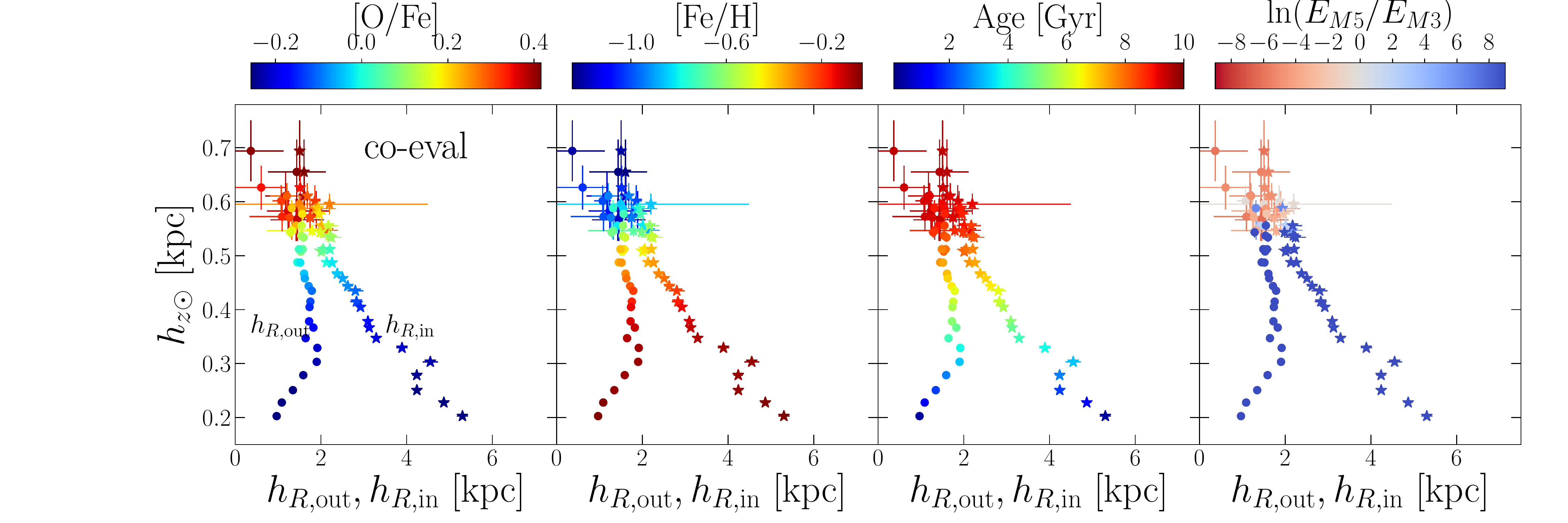}
  \caption{Best fit values of the outer and inner radial scale lengths
    $h_{R,\mathrm{out}}$, $h_{R,\mathrm{in}}$ vs vertical scale height
    at the Solar position $h_{z\odot}$ for mono-age populations
    obtained fitting a single flaring exponential model for
    $\zeta(z|R)$ and a broken exponential for $\Sigma(R)$ -- model {\bf
      M5} to the simulation {\bf FB80}. The data selection, model fit
    and colour-coding are the same as that of
    Fig.~\ref{fig:FB10_coeval_single_exp_hR_hz_I}. The small values of
    $h_{z\odot}$, in comparison to
    Fig.~\ref{fig:FB10_coeval_broken_exp_hR_hz}, indicate the absence
    of a thick disc. Young, $\alpha$-poor and metal-rich populations
    have sharp breaks, i.e. significant differences
    ${h_{R,\mathrm{in}} -h_{R,\mathrm{out}}}$, while old,
    $\alpha$-rich, metal-poor populations have
    $h_{R,\mathrm{out}} \approx h_{R,\mathrm{in}}$, and thus flat
    exponential profiles. The right panel is colour-coded by the Bayes
    factor comparing the single versus broken exponential models for
    $\Sigma(R)$. High-[O/Fe] populations are better described by
    single exponentials (red points), while low [O/Fe] populations
    strongly favour the broken exponential model (blue points).}
  \label{fig:FB80_coeval_broken_exp_hR_hz}
\end{figure*}

Fig.~\ref{fig:FB80_coeval_broken_exp_hR_hz}, to be compared to
Fig.~\ref{fig:FB10_coeval_broken_exp_hR_hz}, shows the best fit values
of $h_{R,\mathrm{in}}$ and $h_{R,\mathrm{out}}$ obtained fitting the
broken exponential model to the radial profiles of co-eval populations
in the simulation {\bf FB80}. Once more, the scale height and scale
length of old populations are typically small in comparison to the
clumpy simulation {\bf FB10}. On the other hand, young populations
have similar parameter values in both simulations.

\begin{figure}
  \includegraphics[width=\columnwidth]{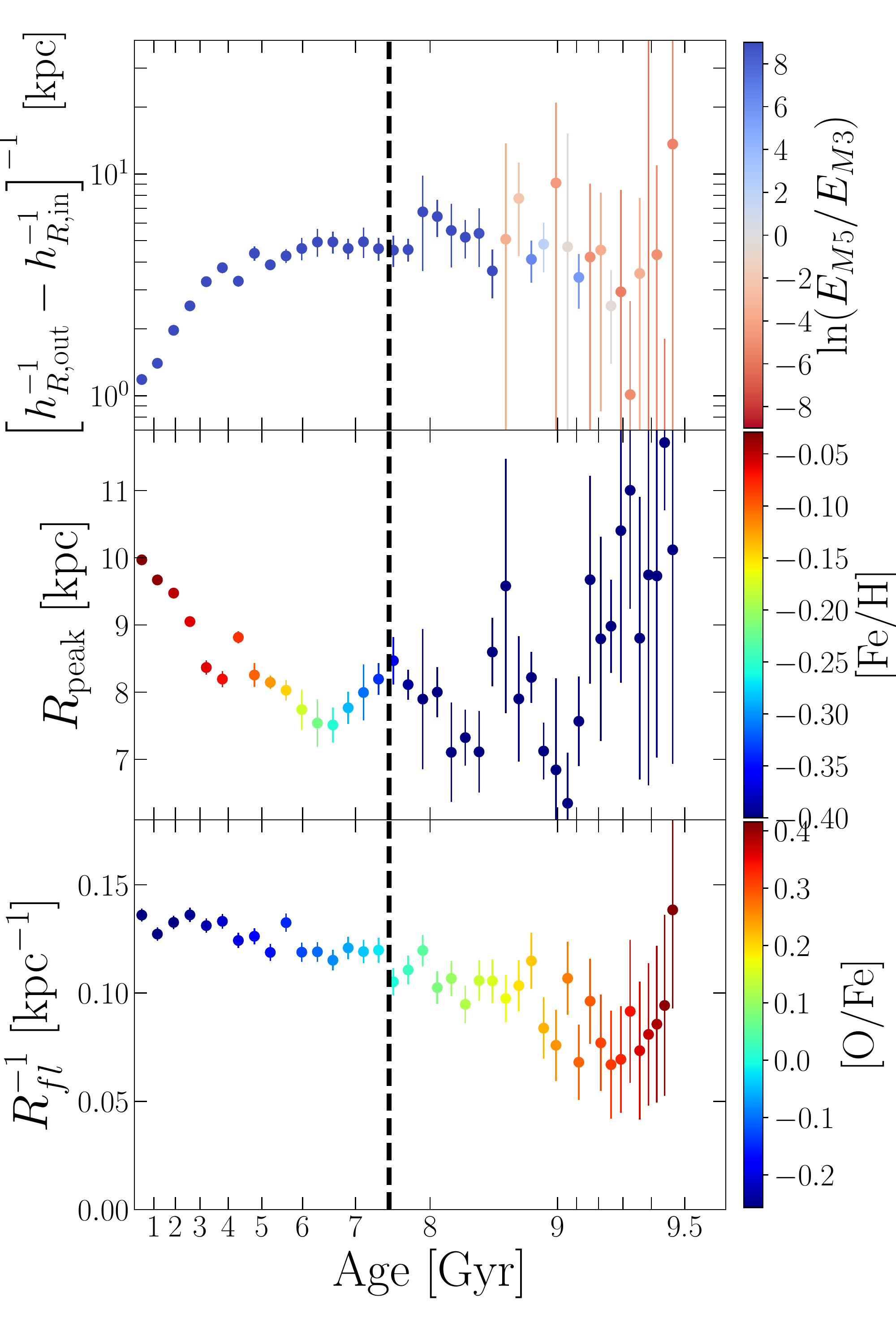}
  \caption{The same quantities shown in
    Fig.~\ref{fig:FB10_coeval_broken_exp_Rpeak_Rfl}, but now for the
    non-clumpy simulation {\bf FB80}. The vertical dashed line crosses
    the bottom panel at [O/Fe]=0 (corresponding now to an age
    $\approx 7.5$ Gyr): low-$\alpha$ populations are again described
    by a broken $\Sigma(R)$, with peak sharpness $\Delta$ decreasing
    with age, peak position decreasing with age for ages $\lesssim 4$
    Gyr and presenting some wiggles for larger ages, and flaring level
    $R_{fl}^{-1}$ generally larger than in {\bf FB10}. High-$\alpha$
    populations have $\Sigma(R)$ better described by single
    exponentials, thus with large uncertainties for the over-fitting
    quantities $\Delta$ and $R_\mathrm{peak}$.}
  \label{fig:FB80_coeval_broken_exp_Rpeak_Rfl}
\end{figure}

Fig.~\ref{fig:FB80_coeval_broken_exp_Rpeak_Rfl} shows, as a function
of age, the quantities characterizing the break in the radial profiles
and the flaring of the vertical profiles produced by the simulation
{\bf FB80}, and is to be compared to
Fig. \ref{fig:FB10_coeval_broken_exp_Rpeak_Rfl} produced by the
simulation {\bf FB10}. Here, the vertical line defining low and
high-$\alpha$ regions ([O/Fe] $\lessgtr 0$) is at $\approx 7.5$
Gyr. Overall these plots are similar to those for {\bf FB10},
particularly, in the top panel, the quantity $\Delta$,
Eq.~(\ref{eq:delta}), measuring how sharp is the break: young
populations have sharp breaks, but 2 or 3 Gyrs of radial migration
(produced by the spiral arms and equally present in the two
simulations) flattens the radial profiles. In the middle panel, we see
that the position of the break seems to vary more with age
($\approx 2.5$ Gyr) in {\bf FB80}, but this variation is still smaller
than those reported by \cite{2017MNRAS.471.3057M} for the Milky
Way. Finally, the bottom panel shows that in {\bf FB80} the young,
$\alpha$-poor populations seem to flare more than in {\bf FB10}.

\bsp	
\label{lastpage}
\end{document}